\documentclass[aps,preprint,superscriptaddress,nofootinbib]{revtex4-1}

\usepackage{amsmath,amssymb,amsfonts,bm,slashed,graphicx,multirow,color,array,mathtools}
\usepackage{multirow, hyperref,comment}

\newcount\notes
\ifnum\notes=1

\newcommand{\SHn}[1]{{\color{red} [SH:~#1]}}
\else
\newcommand{\SHn}[1]{}
\fi

\newcommand{\lqcd}{\ensuremath{\Lambda_\text{QCD}}}

\newcommand{\Vqd}{\ensuremath{V_{qd}}}

\newcommand{\Vqb}{\ensuremath{V_{qb}}}
\newcommand{\Vud}{\ensuremath{V_{ud}}}

\newcommand{\Vub}{\ensuremath{V_{ub}}}

\newcommand{\Vtd}{\ensuremath{V_{td}}}

\newcommand{\Vtb}{\ensuremath{V_{tb}}}

\newcommand{\cO}{\mathcal{O}}

\newcommand{\cA}{\mathcal{A}}

\newcommand{\bra}[1]{\langle #1|}
\newcommand{\ket}[1]{|#1\rangle}

\newcommand{\nn}{\nonumber}

\usepackage[dvipsnames]{xcolor}
\definecolor{red}{rgb}{0.9, 0,0}

\def\beq{\begin{equation}}
\def\eeq{\end{equation}}
\def\beqa{\begin{eqnarray}}
\def\eeqa{\end{eqnarray}}
\def\ben{\begin{enumerate}}
\def\een{\end{enumerate}}

\makeatletter
\g@addto@macro\bfseries{\boldmath}
\makeatother

\def\Oxford{Rudolf Peierls Centre for Theoretical Physics, Department of Physics, University of Oxford, Oxford OX1 3PU, UK}

\begin{document}

%%%%%%%%%%%%%%%%%%%%%%%%%%%%%%%%%%%%%%%%%
\date{\today}

\title{An Exploration of Higher Order Flavor Sum Rules}
%%%%%%%%%%%%%%%%%%%%%%%%%%%%%%%%%%%%%%%%%

%%%%%%%%%%%%%%%%%%%%%%%%%%%%%%%%%%%%%%%%%

\author{Saquib Hassan}
\email[Electronic address: ]{saquib.hassan@physics.ox.ac.uk} 
\affiliation{\Oxford}

%%%%%%%%%%%%%%%%%%%%%%%%%%%%%%%%%%%%%%%%%

%%%%%%%%%%%%%%%%%%%%%%%%%%%%%%%%%%%%%%%%%
\begin{abstract}
We explore the idea of higher-order flavor sum rules, i.e. sum rules that hold to higher orders in the flavor symmetry-breaking parameters. In particular, we consider isospin sum rules based on $B \to n\pi$ decays, and U-spin sum rules based on $B_d$ and $B_s$ decays to $n$ charged pions and/or kaons. We also demonstrate that sum rules exist to arbitrary orders by considering decay modes with increasing number of final state mesons $n$. Finally, we identify various theoretical and practical issues that need to be addressed in utilizing these sum rules.
\end{abstract}
%%%%%%%%%%%%%%%%%%%%%%%%%%%%%%%%%%%%%%%%%

%%%%%%%%%%%%%%%%%%%%%%%%%%%%%%%%%%%%%%%%%
\preprint{CERN-TH-2019-XXX}
\maketitle
%%%%%%%%%%%%%%%%%%%%%%%%%%%%%%%%%%%%%%%%%

\tableofcontents

\pagebreak

%%%%%%%%%%%%%%%%%%%%%%%%%%%%%%%%%%%%%%%%%
\section{Introduction}
\label{sec:intro}
%%%%%%%%%%%%%%%%%%%%%%%%%%%%%%%%%%%%%%%%%

If flavor symmetries are exact, the amplitudes of some hadronic processes are no longer independent. The equations demonstrating these linear dependences are known as flavor sum rules on the one hand. On the other hand, flavor symmetries are only approximate, so the sum rules cannot be exact relations. Nonetheless, the sum rules may still remain useful depending on the degree of precision required in their applications. For example, the Gronau-London method \cite{Gronau:1990ka} of determining the Cabbibo-Kobayashi-Maskawa~(CKM) quark mixing phase $\alpha \equiv \text{Arg}[\Vtd^{\phantom{*}}  \Vtb^*/ \Vud^{\phantom{*}}  \Vub^*]$ utilizes isospin sum rules relating the amplitudes in $B \to \pi \pi$ processes. Since the sum rules are broken by $\mathcal{O}(\delta)$ corrections, where $\delta$ is a small parameter associated with isospin breaking, this limits the precision of $\alpha$ to $\mathcal{O}(\delta)$. We study these symmetry breaking effects separately for both isospin and U-spin.

To fully utilize the improving precision of flavor experiments, methods of extracting Standard Model~(SM) flavor parameters should have at least a comparable precision. One possibility is to invent new methods that are based on higher-order sum rules, i.e. sum rules with precisions that are higher-order in the symmetry-breaking parameter. This in turn requires that we understand the various sources and structures of flavor symmetry breaking. For instance, isospin is broken by quark mass differences and the electromagnetic~(EM) interaction, which in turn affects the sum rules in a variety of ways: corrections to the effective Hamiltonian, isotriplet--isosinglet mixing, and kinematic corrections due to hadronic mass differences within a multiplet \cite{grossman20133, gronau2013u, gronau2000u}.

In this paper, we explore various aspects of higher-order sum rules and their applications. We begin in the next section with isospin sum rules in $B \to n\pi$. We demonstrate that sum rules exist to any order if $n$ is sufficiently large, and that the sum rules are unaffected by $\pi-\eta-\eta'$ mixing. In the subsequent section, we move on to U-spin sum rules in $B_d$ and $B_s$ decays to $n$ charged pions and/or kaons, and again demonstrate that sum rules exist to any order for large enough $n$. We then present some of these higher-order U-spin sum rules.

%%%%%%%%%%%%%%%%%%%%%%%%%%%%%%%%%%%%%%%%%%%%%%%%%%%%%%%%%%%%%%%%%%%%%%%%%%%%%%%%%%%%%%%%%%%%%%%%%%%%
%%%%%%%%%%%%%%%%%%%%%%%%%%%%%%%%%%%%%%%%%%%%%%%%%%%%%%%%%%%%%%%%%%%%%%%%%%%%%%%%%%%%%%%%%%%%%%%%%%%%
%%%%%%%%%%%%%%%%%%%%%%%%%%%%%%%%%%%%%%%%%%%%%%%%%%%%%%%%%%%%%%%%%%%%%%%%%%%%%%%%%%%%%%%%%%%%%%%%%%%%
%%%%%%%%%%%%%%%%%%%%%%%%%%%%%%%%%%%%%%%%%%%%%%%%%%%%%%%%%%%%%%%%%%%%%%%%%%%%%%%%%%%%%%%%%%%%%%%%%%%%
%%%%%%%%%%%%%%%%%%%%%%%%%%%%%%%%%%%%%%%%%%%%%%%%%%%%%%%%%%%%%%%%%%%%%%%%%%%%%%%%%%%%%%%%%%%%%%%%%%%%

%%%%%%%%%%%%%%%%%%%%%%%%%%%%%%%%%%%%%%%%%
\section{Isospin sum rules: a hand-waving argument}
\label{sec:IsoHW}
%%%%%%%%%%%%%%%%%%%%%%%%%%%%%%%%%%%%%%%%%

In this section we give an informal explanation of following statement: ``There exists a sum rule in $B$ decay amplitudes that holds to arbitrary order''. 
This statement is valid for both isospin and $U$-spin sum rules, but here we will only illustrate it for the specific case of isospin in $B$ decay to pions. 
That is, we show that there exists a sum rule for $B \to n \pi$ that holds to order $(n-2)$ in the breaking parameter.
For $n=2$, this is the known Gronau-London sum rule~\cite{Gronau:1990ka}, that holds to zeroth order and is broken at first order. 
For $n=3$, there is an amplitude sum rule that holds to first order and is broken only at second order.

In order to prove the above statement we note the following:
\begin{itemize}
\item The physical amplitudes and the isospin amplitudes are related by a change of basis matrix.
\item We do not consider the momentum dependence of the amplitude. This is an important issue, and we discuss it at length later on.
\item We consider the weak Hamiltonian to leading order in the following way. We consider tree level decays and QCD penguins as leading order processes. 
Electroweak penguins are considered first order, as they are suppressed by $\alpha_{\rm EM}$.
\item
The weak tree-level decays we are interested in are $b \to d \bar u u$ and $b\to d \bar d d$.
They generate only $\Delta I=1/2$ and $\Delta I=3/2$ transitions.
\item
Isospin breaking effects are always $\Delta I=1$. The reason for this is that the breaking we deal with are mass effects that come as $q\bar q$ and thus are $I=1$ (as they are made of two $I=1/2$ quarks). Other breaking effects due to photon or $Z$ effects are also $I=1$ as the photon and the $Z$ couples to $q \bar q$ with different strength.
\end{itemize}

Next, we note that amplitudes in the isospin basis, $ A^I_i$, are simply linear combinations of amplitudes in the physical basis, $ A^P_i$, in the sense that they are related by:
\beq 
	\label{eq:AIAP}
	A^I_i
=	M_{ij} A^P_j \, ,
\eeq
where $M_{ij}$ is a change of basis matrix containing group theoretical numerical factors, and the index $j$ labels the processes that occur. Therefore, a single zero in the left hand side of Eq.~\eqref{eq:AIAP} above, i.e.~a vanishing isospin amplitude, induces a linear combination of physical amplitudes to vanish. This is a sum rule.
In this case, Eq.~\eqref{eq:AIAP} reduces to
\beq
	\sum_i a^i_{CG} A^P_i
=	0 \, ,
\eeq
such that $a^i_{CG}$ are Clebsch-Gordan coefficients appearing with the physical amplitudes $ A^P_i$.
While this method is not the only way to generate a sum rule, it is the one most relevant to our study. 

Next, we move to the specific case of $B \to n \pi$. 
The decay amplitude from initial state with $I=1/2$ to the $n\pi$ final state, with $I=I_f$, mediated by the effective Hamiltonian $H_{\Delta I}$ can be defined as
\beq
	A_{\Delta I}
=	\langle 1/2|H_{\Delta I}|I_f\rangle \, .
\eeq
%
%such that the initial state is $I=1/2$ and $I_f$ is the final isospin of the state contains $n$ pions. 
Note that in the symmetric case ${\Delta I}$ is enough to define the amplitude.
We note that there are $n+1$ physical decay rates, and thus the same number of amplitudes. 
In the physical basis the counting is easy. 
As a simple illustration, consider the special case of $n=3$: 
here, the only relevant decay modes of the isospin doublet $(B^+, B^0)$ are $B^+ \rightarrow \pi^+ \pi^+ \pi^-$, $\pi^+ \pi^0 \pi^0$ and $B^0 \rightarrow \pi^+ \pi^- \pi^0$, $\pi^0 \pi^0 \pi^0$, i.e. $3+1=4$ decay processes. 
A similar analysis for higher values of $n$ leads one to conclude that there are $n+1$ physical amplitudes (see Table I for additional details). 
In the isospin basis, we note that for even $n$, the final state isospin $I_f$, must be even and for odd $n$, it must be odd. 
For any amplitude that is non zero, and has a final isospin $I_f$, there are two ways to get there from the initial $I=1/2$ $B$ meson, that is with $\Delta I=I_f \pm 1/2$. 
For final state with $I_f=0$ there is only the $\Delta I=1/2$ possibility. Thus we see that there are $n+1$ amplitudes in both cases.
We conclude that $\Delta I$ of the $n+1$ amplitudes are given by 
\beq
	\Delta I 
=	{1 \over 2},~{3 \over 2},\, \dots ,{2n+1 \over 2}.
\eeq
Since the zeroth order effective Hamiltonian can only give rise to a maximum of $\Delta I = 3/2$, the highest $\Delta I$ requires $n-1$ insertions of isospin-breaking operators. This means that at order $(n-2)$ in isospin breaking, the effective Hamiltonian cannot generate the highest $\Delta I$, resulting in one isospin amplitude that is zero. This is basically our proof. We see that for a final state with $n$
pions there is an isospin amplitude that is zero to order $(n-2)$ and
thus generates a sum rule. 

Expanding upon the above discussion, we consider a generic formalism for $SU(2)$ sum-rules for $n$-body final state particles, 
extending the work of \cite{Gronau:1990ka,Grossman:2013lya}.
Let us assume that the system has $SU(2)$ symmetry (whether it be $U$-spin or isospin), which is broken by a set of spurions. 
The matrix element of the decay amplitude $B_\mu \to P_{\alpha_1} ... P_{\alpha_n}$ is defined as
\begin{align}
	\label{eq:Adef}
	A_{\mu \to \alpha_1 ... \alpha_n} 
	\equiv 
	\bra{B_\mu} H \ket{P_{\alpha_1} ... P_{\alpha_n}}  \, , 
\end{align}
where $H$ is the effective Hamiltonian. 
The Wigner-Eckart theorem implies that the amplitude can be written as
\begin{align}
	\label{eq:AXI}
	A_{\mu\to\alpha\beta} = \sum_w X_w \partial_{P_{\alpha_1} ... P_{\alpha_n} B_\mu} I_w \, ,
\end{align}
where $X_w$ are the reduced matrix elements and $I_w$ are group theoretical invariants, formed from the effective Hamiltonian, initial, and final state tensors. An amplitudes sum rule will be 
\begin{align}
	S^{\alpha_1 ... \alpha_n\mu}A_{\mu\to\alpha_1 ... \alpha_n} 
=& 	S^{\alpha_1 ... \alpha_n\mu} \sum_w X_w \partial_{P_{\alpha_1}... P_{\alpha_n} B_\mu} I_w  \nonumber\\
= &	S^{\alpha\alpha_1 ... \alpha_n} \sum_w X_w \sum_q \chi^q_{w,\alpha_1 ... \alpha_n\mu}\exp(i \sigma^q_w)   = 0 \, ,
\end{align}
where $\sigma^q_w$ are the strong phases and $\chi^q_{w,\alpha_1 ... \alpha_n\mu}$ contain the weak phase as well group theoretical coefficients.  

Next, let us assume that the initial states, final states and the effective Hamiltonian have the following $SU(2)$ representations 
\begin{align}
	B_\mu \sim r_i \, , \qquad
	P_\alpha \sim r_f \, , \qquad
	H_i \sim r_H \,  ,
\end{align}
and the symmetry is broken by the spurions $m^B$ with the representation 
\begin{align}
	m^B_\beta \sim r_b \, .
\end{align}
Therefore, the $I_w$ are constructed from all the invariants of the initial/final states, the Hamiltonian and the spurions.
This can by done order by order in the symmetry breaking spurions. To order $p$, the decomposition is
\begin{align}
	\sum_w I_w
=	r_i \otimes r_H \otimes \underbrace{\left( r_f \otimes \dots \otimes r_f \right)}_{n} \otimes \underbrace{\left(  r_b \otimes \dots \otimes r_b \right)}_p  \, ,
\end{align}
where we assume $n$ final state and $p$ insertions of the breaking spurion.

We can construct the $I_w$ order by order in the spurions. Consider, for instance, the explicit example of the decay of a $B$ meson into two $K$ and $\pi$ mesons. 
The relevant $U$-spin doublets are given by $B=(B_d,B_s); M^-=(\pi^-,K^-); M^+=(K^+,-\pi^+)$. 
Note that $M^-$ is conjugate to $M^+$ since these are $SU(2)$ doublets (recall that $\epsilon_{ab}$ is an invariant symbol under the $SU(2)$ group action). 
Charge conservation requires the final state to be obtainable from a direct product of $M^+$ and $M^-$, in which case we have $\mathbf{2}\otimes \mathbf{2}=\mathbf{3}\oplus \mathbf{1}$. 
Generalizing further, in the case of four meson final states, we have: $\mathbf{2}\otimes \mathbf{2}\otimes \mathbf{2}\otimes \mathbf{2}=\mathbf{5}\oplus \mathbf{3} \oplus \mathbf{3} \oplus \mathbf{3} \oplus \mathbf{1} \oplus \mathbf{1}$.
Observing this pattern, the tensor product of $n$ such doublets (where $n$ is even) is given by the following formula:
\beq
	\label{eq:2n}
	\bigotimes^n \mathbf{2}
=	\bigoplus_{k=0}^{n/2}\frac{n!(n-2k+1)}{k!(n-k+1)!}(\mathbf{n-2k+1}).
\eeq
At this stage, one can see that the spurion, being a triplet, can be readily multiplied as follows:
\beq
	\label{eq:2n3}
	\mathbf{3}\otimes \bigotimes^n \mathbf{2}
 = 	\bigoplus_{k=0}^{n/2}\frac{n!(n-2k+1)}{k!(n-k+1)!}((\mathbf{n-2k+3})\oplus(\mathbf{n-2k+1}) \oplus (\mathbf{n-2k-1})),
\eeq
and more factors of spurion triplets can be multiplied to the above for higher order breaking effects. 
Since for $U$-spin, $r_H =r_i = \mathbf{2}$, their effects can be added to to above formula by just taking $n \to n+2$ in Eqs.~\eqref{eq:2n}--\eqref{eq:2n3}. 
We will use this expression to explicitly count sum rules in Section~\ref{sec:Uspin} below.

%%%%%%%%%%%%%%%%%%%%%%%%%%%%%%%%%%%%%%%%%
\section{Higher order Isospin sum rules in $B \to n\pi$ decays}
\label{sec:IsoBnpi}
%%%%%%%%%%%%%%%%%%%%%%%%%%%%%%%%%%%%%%%%%

In this section, we present a more rigourous version of the proof discussed above.
As mentioned in the introduction, isospin breaking is generated by $u$--$d$ quark mass splitting and their different electromagnetic charges. 
This breaking may manifest itself as corrections to the effective Hamiltonian, as $\pi$--$\eta$--$\eta'$ mixing, and as kinematic corrections. 
For the former two, one may parametrize their effects by $\delta \sim \cO\left( \tfrac{m_u - m_d}{\lqcd}, \alpha_{\text{EM}} \right)$, while for the kinematic corrections, we argue later that the appropriate parameter is $\sim \cO\left( \tfrac{\lqcd}{m_B/n} \right) \cdot \delta $ in most of the phase space. 
Several studies have attempted to determine more carefully the size of the first-order isospin breaking effects~\cite{Gardner:1998gz,Gronau:2005pq}, estimating $\delta$ to be about $2\%$~\cite{Gronau:2005pq}.

We assume the limit of infinite $b$ quark masses, so we can ignore effects of the kinematic corrections. 
With this assumption, we show that sum rules exist to any order in $\delta$ given a sufficient number of final state mesons. 
\begin{comment}
We then discuss the possibility of using these higher-order sum rules to extract $\alpha \equiv \text{Arg}[\Vtd^{\phantom{*}}  \Vtb^*/ \Vud^{\phantom{*}}  \Vub^*]$.\
\end{comment}

%%%%%%%%%%%%%%%%%%%%%%%%%%%%%%%%%%%%%%%%%
\subsection{Proving the existence of higher-order isospin sum rules}
%%%%%%%%%%%%%%%%%%%%%%%%%%%%%%%%%%%%%%%%%

We begin with corrections to the Hamiltonian, which can be parametrized using the spurion approach. $B \to n\pi$ decays are generated at parton level by $b \to q\bar{q}d$ transitions, with $q = u$, $d$.  In the SM, at leading order in the Fermi coupling $G_F$, the weak Hamiltonian $H_0$ responsible for such transitions furnish the $\mathbf{2}$ and $\mathbf{4}$ of isospin with $I_3 = \tfrac{1}{2}$, i.e. they transform as tensor operators $T^{(1/2)}_{1/2}$ or $T^{(3/2)}_{1/2}$ using the standard notation of $T^{(I)}_{I_3}$. The former receive contribution from both tree-level and QCD penguin diagrams, whereas the latter only receive tree-level contributions. Isospin breaking from the $u$--$d$ quark mass splitting, or from their electromagnetic charge difference, is encoded by a $T^{(1)}_{0}$ isospin breaking spurion, $M_\delta$. In the $u = (1,0)$, $d = (0,1)$ basis, the $B$ mesons furnish an isospin antidoublet $B = (B^+, B^0)$, and the pion triplet and spurion have the form
\begin{equation}
	\Pi = \begin{pmatrix} \pi^0/\sqrt{2} & \pi^+ \\ \pi^- & -\pi^0/\sqrt{2} \end{pmatrix}\,, \qquad M_\delta = \delta\begin{pmatrix} 1 & 0 \\ 0 & -1 \end{pmatrix}\,. \label{eq:isospin-pion-defn}
\end{equation}
Note that off-diagonal $M_\delta$ entries are forbidden by electromagnetic charge conservation. 
Tensor products of $k$ insertions of this spurion operator into the Hamiltonian encodes isospin breaking to $\mathcal{O}(\delta^k)$, including effects of electroweak penguins.

In the $B \to n\pi$ phase space, we label each momentum consistently under an energy ordering, i.e. the momenta are assigned to an ordered set $\{p_1, p_2,\ldots,p_n\}$ such that $E_1 \le E_2 \le \ldots \le E_n$. In the case of degeneracy, the degenerate momenta are then ordered by their momenta components, e.g. beginning with the $x$-component.
The amplitudes are then written in energy-ordered form, $A_{c_1 \ldots c_n}$, in which $c_i \in \lbrace +,-,0 \rbrace$ is the charge of the pion with momentum $p_i$, and the parent $B$ is identified by charge conservation. Under this convention, e.g.~the energy-ordered final state $\{\pi^+(p_1), \pi^-(p_2),\ldots\}$ is distinct from $\{\pi^-(p_1), \pi^+(p_2),\ldots\}$, with amplitudes $\cA_{+-\ldots}$ and $\cA_{-+\ldots}$ respectively, and we have restricted the phase space to an energy-ordered wedge with hypervolume $(1/n!)$th of that of the full phase space.

The starting Hamiltonian can be written in the isospin-decomposed form $H_0 = H_{\frac12} + H_{\frac32}$, and we denote by $(M_\delta)^{\le k}$ the tensor products of up to $k$ insertions of the isospin breaking spurion. 
Then, $(M_\delta)^{\le k} H_0$ transforms as $T^{(1/2)}_{1/2} \oplus T^{(3/2)}_{1/2} \oplus \ldots \oplus T^{(k+3/2)}_{1/2}$.
We have only kept one $T^{(I)}_{1/2}$ representation for each $I$, since we are only looking at $I_3=\tfrac{1}{2}$ components (we can always regard the linear combination of multiple $T^{(I)}_{1/2}$ of the same $I$ as the $I_3=\tfrac{1}{2}$ component of a single representation). Note also that the highest-weight tensor representation $T^{(k+3/2)}_{1/2}$ involves only $H_{\frac32}$ and not $H_{\frac12}$. Now acting this on $|B\rangle$, we find that
\begin{equation}
 (M_\delta)^{\le k} H_0 \big| B\big\rangle \sim |0\rangle \oplus 2\Big[\! |1\rangle\! \oplus |2\rangle \oplus  \ldots \oplus \! |k+1\rangle\Big]\!\oplus |k+2\rangle\,,
\end{equation}
Only the $I_3 = 1$ and $0$ components are non-zero, being furnished by $B^+$ and $B^0$ respectively. There are two linearly independent occurrences of each $|I\rangle$ representation, with the exception of the singlet and the highest weight representation, which are unique. Again the highest weight $|k+2\rangle$ involves only $H_{\frac32}$.

Next we consider the $n\pi$ final states. We first consider specific points in phase space called the \emph{symmetric points}. These are points where $A_{c_1,\ldots,c_n}$ is unchanged under all possible permutations of the same set of charges $\lbrace c_1,\ldots,c_n \rbrace$. We defer further discussion about the symmetric point to Appendix~\ref{app:symmetric-point}. In that case, these $n\pi$ final states project only to the totally symmetric isospin states, i.e. $\langle n \pi | \sim \langle n | \oplus \langle n-2| \oplus \ldots$, with the lowest-weight representation $\langle 0|$ for even $n$ and $\langle 1 |$ for odd $n$.

We define $\mathcal{O}(\delta^k)$ amplitude as amplitudes that include up to and including $\mathcal{O}(\delta^k)$ corrections (i.e. from $(M_\delta)^{\le k} H_0$), and hence have a precision of $\mathcal{O}(\delta^{k+1})$. 
One can now count the number of independent $\mathcal{O}(\delta^k)$ isospin amplitudes (equivalent to Wigner-Eckart reduced matrix elements), for the case of even and odd $n$ and $k$. At the same time, we can count the number of neutral and charged $B \to n\pi$ decay modes at the symmetric point. Since charge permutation does not matter at the symmetric point, we are only interested in distinct charge compositions, so this corresponds to the number of solutions of the Diophantine equations $2 a + b = n$ and $2a + b + 1 = n$ respectively -- the number of ways the $n\pi$ final state may contain $a$ $\pi^-$'s and $b$ $\pi^0$'s -- with the requirements $a$, $b \ge 0$. The number of neutral and charged modes for $n$ even and odd are also shown in Table~\ref{tab:IDA}.

\begin{table}[t]
\setlength{\tabcolsep}{10pt}
\begin{tabular}{|c|c|c|c|c|c|}\hline
		                    &\multicolumn{3}{c|}{Isospin amplitudes}                             &\multicolumn{2}{c|}{Physical decay amplitudes} \\
	\hline
	$k$                   &\multicolumn{2}{c|}{$n$}          &N[$\mathcal{O}(\delta^k)$ ampl.] &Neutral                      &Charged\\
	\hline\hline
	\multirow{4}{*}{even} &\multirow{2}{*}{even} &$n\ge k+2$ &$(k+2)^*$                        &\multirow{2}{*}{$n/2+1$}   &\multirow{2}{*}{$n/2$}\\
	\cline{3-4}
	                      &                      &$n\le k$   &$n+1$                            &                             &\\
	\cline{2-6}				
			                  &\multirow{2}{*}{odd}  &$n\ge k+3$ &$k+2$                            &\multirow{2}{*}{$(n+1)/2$} &\multirow{2}{*}{$(n+1)/2$}\\
	\cline{3-4}		                  
			                  &                      &$n\le k+1$ &$n+1$                            &                             &\\          
	\hline				
	\multirow{4}{*}{odd} 	&\multirow{2}{*}{even} &$n\ge k+3$ &$k+2$                            &\multirow{2}{*}{$n/2+1$}   &\multirow{2}{*}{$n/2$}\\
	\cline{3-4}
					              &                      &$n\le k+1$ &$n+1$                            &                             &\\	
  \cline{2-6}            
					             	&\multirow{2}{*}{odd}  &$n\ge k+2$ &$(k+2)^*$                        &\multirow{2}{*}{$(n+1)/2$} &\multirow{2}{*}{$(n+1)/2$}\\
	\cline{3-4}
					              &                      &$n\le k$   &$n+1$                            &                             &\\	
	\hline
\end{tabular}
	\caption{Maximum number of $\mathcal{O}(\delta^k)$ $B \to n\pi$ isospin amplitudes and number of $B \to n\pi$ charged and neutral decay amplitudes, at the symmetric point in phase space. When labelled with an asterisk (*), this indicates that the highest-weight isospin amplitudes only involves $H_{\frac32}$ and not $H_{\frac12}$.}
	\label{tab:IDA}
\end{table}

The Clebsch-Gordan coefficients relating $|I\rangle$ to $(M_\delta)^{\le k} H_0|B\rangle$, and $\langle I|$ to $\langle n\pi|$, specifies the relation between the isospin and the physical decay amplitudes. In other words,
\begin{equation}
	A_{c_1^i \ldots c_n^i} = \sum_{j=1}^{N_I} C_{ij} A^I_j\,,
\end{equation}
where $i$ runs from $1$ to $N_{\text{phys}}$, $N_{\text{phys}}$ and $N_I$ counts the number of decay modes and isospin amplitudes respectively, and $C_{ij}$ comprises products of known Clebsch-Gordon coefficients. This expression is true at any point in phase space; however as we have mentioned before, $N_{\text{phys}}$ only counts distinct charge compositions at the symmetric point. In that case, regardless of whether $n$ is even or odd, we find that $N_{\text{phys}}=n+1$. Meanwhile, from Table~$\ref{tab:IDA}$, when $n \ge k+2$, we find that $N_I = k+2 < n+1$, so the kernel dimension of the mapping $C_{ij}$ is nonzero. This guarantees the existence of $\mathcal{O}(\delta^k)$ sum rules: linear combinations $\sum_i a^{i}_{CG} A_{n\pi}^i = 0$, with a precision of $\mathcal{O}(\delta^{k+1})$. In contrast, when $n \le k+1$, we find that $N_I = n+1 = N_{\text{phys}}$, so the existence of a sum rule is not guaranteed, demonstrating our claim that we can obtain sum rules to any order in $\delta$ as long as we consider sufficiently large $n$.

At this point, it is useful to introduce the following notation: in $b \to q\bar{q} d$ transitions, CKM unitarity ensures that one may always write the decay amplitudes in the factorized form
\begin{equation}
	\label{eqn:AFTUD}
	A_{c_1^i \ldots c_n^i} = A_{c_1^i \ldots c_n^i}^t\lambda_t +  A_{c_1^i \ldots c_n^i}^u\lambda_u\,,\qquad \lambda_q \equiv \Vqb^*\Vqd
\end{equation}
such that $A_{c_1^i \ldots c_n^i}^t$ is exclusively produced by penguin operators, while $A_{c_1^i \ldots c_n^i}^u$ may have both tree and penguin contributions, and both $A_{c_1^i \ldots c_n^i}^t$ and $A_{c_1^i \ldots c_n^i}^u$ contain only strong phases. (We emphasize that the $t$ subscript is for notational convenience: In practice the penguin contributions below $m_B$ are generated by charm and up-quark loops most dominantly, not by top quarks.) In the cases indicated by an asterisk (*) in Table~$\ref{tab:IDA}$, the highest-weight isospin amplitudes only involves $H_{\frac32}$ and not $H_{\frac12}$, and hence has a CKM structure $\lambda_u$. This implies that $A_{c_1^i \ldots c_n^i}^t$ cannot project to this isospin amplitude, and therefore there will be an additional sum rule for $A_{c_1^i \ldots c_n^i}^t$.

\subsection{Specific examples}

For example, for $n=2$ and $k=0$, the full decay amplitude relation and the $A_f^t$ amplitude relation are explicitly
\begin{equation}
	\label{eqn:2PSR}
	\sqrt{2}A_{+0} - A_{+-} + A_{00} = 0\,,\qquad A^t_{+-} - A^t_{00} = 0\,.
\end{equation}
We note that the first sum rule differs from the one used in \cite{Gronau:1990ka} due to differences in convention. See App.~\ref{app:isospin-sum-rules-convention} for details.

For $n=3$, the amplitude relations are
\begin{gather}
	A_{++-} - 2A_{+00} + 2\sqrt{2} A_{+-0} - \frac{2\sqrt{2}}{3}A_{000} = 0, \quad \text{for }k=0,1 \nn\\
	\begin{cases} A_{+-0} - \frac{1}{3} A_{000} = 0, \quad \text{for }k=0 
	\\ A^t_{+-0} - \frac{1}{3} A^t_{000} = 0, \quad \text{for }k=1 \end{cases} \label{eqn:3PSR}
\end{gather}

For $n=4$, the amplitude relations are
\begin{gather}
  A_{++-0} - \frac{2}{3} A_{+000} -\frac{\sqrt{2}}{4} A_{++--} + \sqrt{2} A_{+-00} - \frac{\sqrt{2}}{6}A_{0000} = 0, \quad \text{for } k=0,1,2 \nn\\
  \begin{cases} A_{++--} - 4 A_{+-00} +\frac{2}{3}A_{0000} = 0, \quad \text{for } k=0,1 \\
  A^t_{++--} - 4 A^t_{+-00} +\frac{2}{3}A^t_{0000} = 0, \quad \text{for } k=2  \end{cases} \nn\\
  A_{+000} - \frac{3\sqrt{2}}{2} A_{+-00} + \frac{\sqrt{2}}{2}A_{0000} = 0, \quad \text{for }k=0 \nn\\
  A^t_{+-00} - \frac{1}{3}A^t_{0000} = 0, \quad \text{for }k=0
  \label{eqn:4PSR}
\end{gather}

\subsection{Moving away from the symmetric point} \label{sec:away-from-symmetric-point}
The above discussion applies to symmetric points in phase space. For more general points, the sum rules above still apply if we replace the amplitudes above by the symmetrized amplitudes, i.e.
\begin{equation}
\mathcal{A}_{c_1,\ldots, c_n}  \equiv \frac{1}{s(c_1,\ldots, c_n)}\sum_{\sigma \in S_n } A_{\sigma(c_1,\ldots,c_n)}\,,
\end{equation}
where $\sigma(c_1,\ldots,c_n) \in S_n$ is a permutation of the set of pion charges, and $s(c_1,\ldots,c_n)$ are appropriate definitions of the symmetry factor. (The symmetrized amplitude describes decays to a symmetric linear superposition of all modes related by charge permutation, which again projects to totally symmetric isospin states due to Bose statistics.) However, on top of that, there may be other sum rules that do not involve the totally symmetric $\langle I|$ representations.
To illustrate this, let us consider the example of $n=3$, $k=0$. At the symmetric point, from Table~\ref{tab:IDA}, we have $N_{\text{phys}} = n+1 = 4$ and $N_I = k+2 = 2$, suggesting two sum rules. For more general points in phase space however, since amplitudes now differ under permutations of the charge indices, we have to count distinct permutations for a given charge composition as separate modes, and we find that $N_{\text{phys}} = 13$ instead. Meanwhile, final-state isospin representations that are not totally symmetric now also come into play, so we have
$\langle 3 \pi| \sim \langle 0| \oplus 3 \langle 1| \oplus 2\langle 2| \oplus \langle 3|$, among which there are $1 + 3\times 2 + 2\times 2 + 2= 13$ states with $I_3=0$ or $1$, in agreement with the number of modes. %(Unlike what was done for $(M_\delta)^{\le k} H_0$, here even though we are interested only in the $I_3 = 0$ or $1$ final states, we do not combine multiple representations of the same $I$ into just two independent representations, since we want to be able to invert the isospin representations of the final states back into in original physical basis to get the sum rules.)
For $k=0$, where $(M_\delta)^{\le k} H_0|B\rangle \sim |0\rangle \oplus 2|1\rangle \oplus |2\rangle$ we find that $N_I = 9$, i.e.~only 9 reduced matrix elements. This suggests $13-9=4$ sum rules, two more than at the symmetric point.

\subsection{Effects of $\pi-\eta$ mixing}

So far, we have assumed that the pions are isotriplets. Isotriplet-isosinglet mixing, i.e. mixing of $\pi^0$ with $\eta$ and $\eta'$ in the physical states, may be encoded via a single mixing angle $\theta$, such that $\pi^0 = \pi_{\text{phys}} \cos\theta + \eta_{\text{phys}} \sin\theta $ and $\eta = -\pi_{\text{phys}} \sin\theta + \eta_{\text{phys}} \cos\theta$. Since isotriplet-isosinglet mixing is a manifestation of isospin breaking in the initial and final states, it necessarily originates from the same physics parametrized by $\delta$. That is, formally we expect $\theta \sim \delta$.

At first glance, one might expect $\mathcal{O}(\delta^k)$ sum rules derived using $n \pi$ to include not just decay amplitudes to pure $n \pi_{\text{phys}}$, but also those to $(n-m) \pi_{\text{phys}} (m) \eta_{\text{phys}}$ as well for $0 \le m \le n$, i.e. we cannot simply replace $\pi$ by $\pi_{\text{phys}}$ to get pure $\pi_{\text{phys}}$ sum rules without $\eta_{\text{phys}}$. However, this intuition appears to be wrong: the simple replacement of $\pi$ by $\pi_{\text{phys}}$ gives the correct sum rules.

We first illustrate this with the $n=3$, $k=1$ example at the symmetric point. Since $k=1$ sum rules are supposed to only receive $\mathcal{O}(\delta^2)$ corrections, we just want to check whether mixing introduces $\mathcal{O}(\delta)$ corrections to these sum rules, so we can approximate $\pi^0 \simeq \pi_{\text{phys}} + \eta_{\text{phys}} \theta $. Now plugging this into, say, the first equation in Eq.~(\ref{eqn:3PSR}), we get in terms of the decay amplitudes to physical states
\begin{equation}
\begin{aligned}
	&\left(A^{\text{phys}}_{++-} - 2A^{\text{phys}}_{+00} + 2\sqrt{2} A^{\text{phys}}_{+-0} - \frac{2\sqrt{2}}{3}A^{\text{phys}}_{000}\right)\\
	&-\theta \left[
	2\left( A^{\text{phys}}_{+\eta 0} + A^{\text{phys}}_{+0\eta} \right)
	- 2\sqrt{2} A^{\text{phys}}_{+-\eta}
	+ \frac{2\sqrt{2}}{3}\left( A^{\text{phys}}_{\eta 00} + A^{\text{phys}}_{0\eta 0} + A^{\text{phys}}_{00\eta} \right)
	\right] 	= 0 \,, \label{eq:isospin-pi-eta-mixing-example}
\end{aligned}
\end{equation}
where we have dropped terms of $\mathcal{O}(\theta^2)$ and above. Since we are at the symmetric point, rotational invariance allows us to simplify the terms in the square brackets as
\begin{equation}
	2\sqrt{2} \left( \sqrt{2} A^{\text{phys}}_{+0\eta} - A^{\text{phys}}_{+-\eta} + A^{\text{phys}}_{00\eta} \right)\,.
\end{equation}
We now argue that this expression is actually of size $\mathcal{O}(\delta)$. Up to $\mathcal{O}(\delta)$ corrections, we can treat $\pi_{\text{phys}}$ and $\eta_{\text{phys}}$ as $\pi$ and $\eta$ respectively, and since the addition of a singlet $\eta$ does not affect isospin decomposition of the final state, the two-pion $\mathcal{O}(\delta^0)$ sum rules from Eq.~(\ref{eqn:2PSR}) implies that this expression is of size $\mathcal{O}(\delta)$. Since this expression is further multiplied by $\theta$ in Eq.~(\ref{eq:isospin-pi-eta-mixing-example}), it becomes $\mathcal{O}(\delta^2)$ and can hence be dropped. Therefore, we are left with just
\begin{equation}
A^{\text{phys}}_{++-} - 2A^{\text{phys}}_{+00} + 2\sqrt{2} A^{\text{phys}}_{+-0} - \frac{2\sqrt{2}}{3}A^{\text{phys}}_{000} = 0\,,
\end{equation}
same as if we had simply replaced $\pi$ in the sum rules by $\pi_{\text{phys}}$.

We explicitly verified that this is the case for all the sum rules in Eq.~(\ref{eqn:3PSR}) and Eq.~(\ref{eqn:4PSR}). In other words, all correction terms from the difference between $\pi^0$ and $\pi_{\text{phys}}^0$ turn out to be of order $\mathcal{O}(\delta^{>k})$ and can hence be neglected. This is usually a consequence of sum rules for lower $n$ as seen in the example above, or sum rules for the same $n$ but lower $k$ (e.g. for $n=4$, $k=2$, when replacing $\pi^0$ by $(1-\tfrac{\theta^2}{2})\pi_{\text{phys}}^0 + \theta \eta_{\text{phys}}$, the correction terms associated with $-\tfrac{\theta^2}{2} \pi_{\text{phys}}^0$ can be neglected due to a $n=4$, $k=0$ sum rule).

While we have only shown this for $n=3,4$ and only at the symmetric point, we believe this to be true in general, although we defer a complete proof to future work.

\subsection{Limitations}

First, we have only considered the corrections to the effective Hamiltonian as well as $\pi$--$\eta$--$\eta'$ mixing from isospin breaking. We have not taken into account kinematic corrections due to the mass splitting of the $B$ mesons and the pions from isospin breaking. The sum rules so far involve amplitudes that utilize the same set of pion momenta. However, once we take into account mass splitting, for certain decay modes, the same set of pion momenta may no longer be kinematically allowed. Therefore, were we to use the closest kinematically allowed amplitudes for each mode instead, it is likely that the sum rules will obtain corrections of the form
\begin{equation}
\sum_i \frac{\partial A_{c_1\ldots c_n}}{\partial \vec{p}_i} \cdot \Delta \vec{p}_i \sim \frac{A_{c_1\ldots c_n}}{E_\pi} \cdot \lqcd \cdot \delta \sim  \left(\frac{\lqcd}{m_B/n} \cdot \delta \right) A_{c_1\ldots c_n}. \label{eq:kinematic-correction}
\end{equation}
Here, $\Delta\vec{p}_i$ is the typical difference between the forbidden and closest allowed momenta, which we assume to be $1/n$ the scale of the mass splitting $\lqcd \cdot \delta$, and $E_\pi$ is the typical energy of the pion in the final state, assumed to be of order $m_B/n$. We see that the corrections are typically of order $\tfrac{\lqcd}{m_B/n} \cdot \delta > \delta^2$, so $\alpha$ extracted from $\mathcal{O}(\delta)$ sum rules may actually not reach the expected precision of $\mathcal{O}(\delta^2)$. Also, the above estimate for the kinematic correction size is too small in the vicinity of an intermediate resonance, since then $\tfrac{\partial A_{c_1\ldots c_n}}{\partial \vec{p}_i} \sim \tfrac{A_{c_1\ldots c_n}}{\Gamma}$ instead, where $\Gamma$ is the width of the resonance. For recent work on the effects of resonance on sum rules, see \cite{schacht2021enhancement}. One may also have to consider possible interplay with $\pi-\eta-\eta'$ mixing due to the larger kinematic corrections from the mass difference between $\pi$ and $\eta$. We hope to fully address these kinematic issues in future work.
%%%%%%%%%%%%%%%%%%%%%%%%%%%%%%%%%%%%%%%%%%%%%%%%%%%%%%%%
%%%%%%%%%%%%%%%%%%%%%%%%%%%%%%%%%%%%%%%%%%%%%%%%%%%%%%%%%%%%%%%%%%%%%%%%%%%%%%%%%%%%%%%%%%%%%%%%%%%%
%%%%%%%%%%%%%%%%%%%%%%%%%%%%%%%%%%%%%%%%%%%%%%%%%%%%%%%%%%%%%%%%%%%%%%%%%%%%%%%%%%%%%%%%%%%%%%%%%%%%
%%%%%%%%%%%%%%%%%%%%%%%%%%%%%%%%%%%%%%%%%%%%%%%%%%%%%%%%%%%%%%%%%%%%%%%%%%%%%%%%%%%%%%%%%%%%%%%%%%%%

%%%%%%%%%%%%%%%%%%%%%%%%%%%%%%%%%%%%%%%%%
\section{U-spin sum rules in $B_d$, $B_s$ decay to charged pions/kaons}
\label{sec:Uspin}
%%%%%%%%%%%%%%%%%%%%%%%%%%%%%%%%%%%%%%%%%

We now move on to the topic of U-spin sum rules. In particular, we consider the decays of $B_d$ and $B_s$ to $n$ charged pions and/or kaons, where $n$ is even, i.e. $B_d/B_s \to \tfrac{n}{2}(\pi^+/K^+)\tfrac{n}{2}(\pi^-/K^-)$. U-spin breaking is generated by $d$--$s$ quark mass splitting, and may manifest itself as corrections to the effective Hamiltonian, and as kinematic corrections. The small parameter is now given by $\epsilon \sim \mathcal{O}( \tfrac{m_s - m_d}{\lqcd}) \sim 0.3$ for corrections to the Hamiltonian, and $\sim \left( \tfrac{\lqcd}{m_B/n} \right) \cdot \epsilon$ for kinematic corrections. In this section, we again assume the $m_B \to \infty$ limit and hence ignore kinematic corrections. With this assumption, we prove that sum rules exist to any order in $\epsilon$ given sufficient number of final state mesons. We also present some of the higher-order sum rules below.

\subsection{Sum rules to higher order}

The mesons involved in the decay all furnish U-spin doublets
\begin{equation}
B = \begin{pmatrix} B_d \\ B_s \end{pmatrix}, \quad M ^- = \begin{pmatrix} \pi^- \\ K^- \end{pmatrix}, \quad M^+ =  \begin{pmatrix} K^+ \\ -\pi^+ \end{pmatrix}\,.
\end{equation}
Meanwhile, the weak Hamiltonian can be written as $H_0 = H^t + H^p$, where
\begin{equation}
H^t = \mathcal{T} \begin{pmatrix} V_{ud} V_{ub}^* \\ V_{us} V_{ub}^* \end{pmatrix}, \quad H^p = \mathcal{P} \begin{pmatrix} V_{cd} V_{cb}^* \\ V_{cs} V_{cb}^* \end{pmatrix}\,,
\end{equation}
where $\mathcal{T}$ and $\mathcal{P}$ are complex numbers that contain the strong phases \cite{Grossman:2013lya}. Note that despite the notation, $\mathcal{T}$ may contain penguin contributions with the same CKM structure as tree-level contributions. Since $\mathcal{T}$ and $\mathcal{P}$ are unknown, rather than think of $H_0$ as a linear combination of the $U_3 = \pm \tfrac{1}{2}$ components of a single $\mathbf{2}$ with unknown coefficients, it is more useful here to think of $H_0$ as $H_{\frac12} + H_{-\frac12}$, where $H_{\frac12}$ and $H_{-\frac12}$ are $U_3 = \pm \tfrac{1}{2}$ components of two \emph{inequivalent} $\mathbf{2}$ without coefficients. In other words, the unknown coefficients above can be thought of as causing $H_{\frac12}$ and $H_{-\frac12}$to have different reduced matrix elements; hence the view that they come from inequivalent representations. For additional details on U-spin decomposition (including larger $SU(3)$), see \cite{savage19913, grossman20133, brod2012consistent, grossman2019u, hiller2013s, hiller20133, dery2021probing, Dery:2020lbc} and references therein. In particular, \cite{grossman20133} gives an elaborate account of two methods for extracting $SU(3)$ sum rules without prior need for separate use of $SU(2)$, although such generality will not be necessary for our calculations.

U-spin breaking is encoded by a $T^{(1)}_{0}$ spurion
\begin{equation}
M_\epsilon = \epsilon \begin{pmatrix} 1 & 0 \\ 0 & -1 \end{pmatrix} \,.
\end{equation}
Again, we denote by $(M_\epsilon)^{\le k}$ the tensor products of up to $k$ insertions of the U-spin breaking spurion. Then $(M_\epsilon)^{\le k} H_{\frac12}$ transforms as $T^{(1/2)}_{1/2} \oplus T^{(3/2)}_{1/2} \oplus ... \oplus T^{(k+1/2)}_{1/2}$, and $(M_\epsilon)^{\le k} H_{-\frac12}$ as $T^{(1/2)}_{-1/2} \oplus T^{(3/2)}_{-1/2} \oplus ... \oplus T^{(k+1/2)}_{-1/2}$. Since $T^{(U)}_{1/2}$ and $T^{(U)}_{-1/2}$ belong to inequivalent $\mathbf{2U+1}$ representations, this implies that $\Delta U_3 = \tfrac{1}{2}$ and $\Delta U_3=-\tfrac{1}{2}$ decay amplitudes are related to two disjoint sets of U-spin amplitudes (Wigner-Eckart reduced matrix elements), so sum rules should only relate decay amplitudes of the same $\Delta U_3$. Also, $|\Delta U_3| \ne \tfrac{1}{2}$ decays are explicitly forbidden, unless we go to higher orders in $G_F$, e.g. via box diagrams. (These considerations were absent when discussing isospin sum rules because only $\Delta I_3 = \tfrac{1}{2}$ decay amplitudes in $B \to n\pi$ were allowed by electric charge conservation.)

Therefore, let us restrict our analysis to the $\Delta U_3 = \tfrac{1}{2}$ transitions. We denote $|B_d\rangle$ and $|B_s\rangle$ collectively as $|B\rangle$. In the following paragraphs, we first count the number of U-spin amplitudes, $N_U$, using Eq. (\ref{eq:2n}). We then calculate the number the of physical decay amplitudes, $N_{\text{phys}}$. The Wigner-Eckart theorem allows us to express the physical amplitudes in terms of U-spin amplitudes, along with factors of Clebsch-Gordan coefficients, in accordance with Eq. (\ref{eq:AIAP}), where we may now replace the isospin amplitudes $A^{I}_i$ with U-spin amplitudes $A^{U}_{i}$. Now, $N_{\text{phys}}- N_U$ gives us the number of zeros on the left hand side of Eq. (\ref{eq:AIAP}), which is precisely the number of sum rules.

Acting $(M_\epsilon)^{\le k} H_{\frac12}$ on $|B\rangle$ gives
\begin{equation}
(M_\epsilon)^{\le k} H_{\frac12} \big| B \big\rangle \sim |0\rangle \oplus 2\Big[\! |1\rangle\! \oplus \ldots \oplus \! |k\rangle\Big]\!\oplus |k+1\rangle\,, \label{eq:Uspin-initial-state-decomp}
\end{equation}
where the actual linear combinations involve Clebsch-Gordan coefficients. The right hand side of Eq. (\ref{eq:Uspin-initial-state-decomp}) gives us the irreducible representations, along with their multiplicities, from which we may obtain the number of U-spin amplitudes. Consider the final state comprising $n$ charged pions and/or kaons, at some general point in phase space, where $n$ must be even due to electric charge conservation. We order the final state mesons first by charge, and next by energy. The final state U-spin decomposition is given by
\begin{equation}
\left(\langle \tfrac{1}{2}| \right)^n = \sum_{m=0}^{n/2} \frac{n!(n-2m+1)}{m!(n-m+1)!}\langle \tfrac{n}{2} - m|,
\end{equation}
which upon comparing with Eq.~(\ref{eq:Uspin-initial-state-decomp}), one finds the maximum number of $\Delta U_3=\tfrac{1}{2}$ U-spin amplitudes to be
\begin{equation}
N_U = \begin{dcases}
2\sum_{m=0}^{n/2} \frac{n!(n-2m+1)}{m!(n-m+1)!} - \frac{n!}{(\frac{n}{2})!(\frac{n}{2}+1)!}, &\mbox{ for } k \ge \frac{n}{2}\\
2\sum_{m=\frac{n}{2}-k-1}^{n/2} \frac{n!(n-2m+1)}{m!(n-m+1)!} - \frac{n!}{(\frac{n}{2})!(\frac{n}{2}+1)!} - \frac{n!(2k+3)}{(\frac{n}{2}-k-1)!(\frac{n}{2}+k+2)!}, &\mbox{ for }  0 \le k \le \frac{n}{2}-1\\
\end{dcases}
\label{eq:Uspin-NU}
\end{equation}

Next we count the number of $\Delta U_3 = \tfrac{1}{2}$ decay modes, i.e. the number of physical decay amplitudes, $N_{\text{phys}}$. Since $B_d$ can decay to $U_3 = 1$ states and $B_s$ to $U_3 = 0$ states, the number of modes is simply given by the sum of the number of $U_3 = 1$ and $0$ final states.
\begin{equation}
N_{\text{phys}} = \sum_{m=0}^{n/2}\begin{pmatrix} n/2 \\ m \end{pmatrix}\begin{pmatrix} n/2 \\ m \end{pmatrix} + \sum_{m=1}^{n/2}\begin{pmatrix} n/2 \\ m \end{pmatrix}\begin{pmatrix} n/2 \\ m-1 \end{pmatrix} \,.
\label{eq:Uspin-Nphys}
\end{equation}

From Eq.~(\ref{eq:Uspin-NU}) and Eq.~(\ref{eq:Uspin-Nphys}), we find that despite the seemingly different expression, in fact $N_U = N_{\text{phys}}$ when $n \leq 2k $, so no sum rules are guaranteed to exist. However, once we increase $n$ beyond $2k$, the minimum number of sum rules is given by
\begin{equation}
N_{\text{phys}} - N_U = 2\sum_{m=0}^{\frac{n}{2}-k-1} \frac{n!(n-2m+1)}{m!(n-m+1)!} - \frac{n!(2k+3)}{(\frac{n}{2}-k-1)!(\frac{n}{2}+k+2)!},
\quad \mbox{ for } k \le \frac{n}{2}-1 \,.
\end{equation}
Table~\ref{tab:Uspin-number-sum-rules} lists the minimum number of $\Delta U_3 = \tfrac{1}{2}$ sum rules of precision $\mathcal{O}(\epsilon^{k+1})$ for various $n$ and $k$ values. With this, we complete our argument that sum rules exist to any order in $\epsilon$ provided we include a sufficient number of final state mesons.

\begin{table}[t]
\setlength{\tabcolsep}{10pt}
\begin{tabular}{|c|c|c|c|c|}
\hline
\multirow{2}{*}{$n$} &\multicolumn{4}{c|}{Number of $\Delta U_3 = \tfrac{1}{2}$ sum rules}\\
\cline{2-5}
                     & $k=0$ & $k=1$ & $k=2$ & $k=3$\\
\hline
2                    & 1     & 0     & 0     & 0\\
4                    & 5     & 1     & 0     & 0\\
6                    & 21    & 7     & 1     & 0\\
8                    & 84    & 36    & 9     & 1\\ 
\hline
\end{tabular}
\caption{Minimum number of $\Delta U_3 = \tfrac{1}{2}$ sum rules of precision $\mathcal{O}(\epsilon^{k+1})$, for the decay amplitudes of $B_d$ and $B_s$ to $n$ charged pions and/or kaons.}
\label{tab:Uspin-number-sum-rules}
\end{table}

We can perform the same analysis for $\Delta U_3 = -\tfrac{1}{2}$ decays, in which case we find exactly the same number of U-spin amplitudes, decay modes, and U-spin sum rules as $\Delta U_3 = \tfrac{1}{2}$ decays. It is also useful to count the number of $|\Delta U_3| \ne \tfrac{1}{2}$ that are forbidden at leading order in $G_F$. The maximum number of decay modes for a given charge-specific set of momenta is given by $2^{n+1}$, among which $2 N_{\text{phys}}$ modes are $\Delta U_3 = \pm \tfrac{1}{2}$. Therefore, the number of forbidden modes is simply given by $2^{n+1} - 2N_{\text{phys}}$.

We re-iterate that the discussion above is only based on corrections to the Hamiltonian from U-spin breaking. We have yet to take into account kinematic corrections from the mass splitting in a U-spin multiplet. Therefore, the actual precision of the sum rules may be worse than $\mathcal{O}(\epsilon^{k+1})$ depending on the values of $n$ and $k$. We are also yet to explore how U-spin sum rules may be used in CKM parameter extraction, but we will explore this matter in the future. Some existing ideas based on U-spin symmetry can be found in \cite{Fleischer:1999pa, Fleischer:2007hj} for example.

\subsection{Higher-order U-spin sum rules}

Consider the example mentioned in Sec.~\ref{sec:IsoBnpi} regarding four final state mesons. Restricting ourselves to pieces of the phase space where positively charges final states mesons are written first and negatively charged final state mesons written last, a general amplitude is then of the form $A(B\rightarrow M^+ (p_1) M^+ (p_2) M^- (p_3) M^- (p_4))$ where doublets were defined in Eq. (33). Suppressing momentum labels, this expression is now simply $A_{B\rightarrow M^+ M^+ M^- M^-}$. Additionally, we symmetrize between pairs of positively charged final state mesons and we separately symmetrize between negatively charged final state mesons, if they are different. For instance, $A(B\rightarrow K^+ \pi^+ K^- K^-) \rightarrow (A(B\rightarrow K^+ \pi^+ K^- K^-) + A(B \rightarrow \pi^+  K^+ K^- K^-))/2 = \cA(B\rightarrow K^+ \pi^+ K^- K^-)$. Note that the negative pair does not undergo any further symmetrization since the negatively charged final state mesons are identical particles. If instead there were a $\pi^- K^-$ pair, then we would symmetrize accordingly, but this time there would be a factor of 4 in the denominator: a 2 from the positive pair and another 2 from the negative pair. For channels where pairs of positive and negative mesons are identical there is no symmetrization. For instance, we simply replace $A(B\rightarrow K^+ K^+ \pi^- \pi^-)$ by $\mathcal{A}(B\rightarrow K^+K^+\pi^- \pi^-)$. At the group theoretical level, symmetrizing pairs of final state mesons effectively amounts to considering the tensor product $\mathbf{2}\otimes\mathbf{2}\otimes\mathbf{2}\otimes\mathbf{2}=(\mathbf{1}\otimes \mathbf{3})\otimes(\mathbf{1}\otimes \mathbf{3})$, and extracting the $\mathbf{3}\otimes \mathbf{3}$, i.e. the symmetric representations only. Note however, that the lesser number of U-spin matrix elements will lead to fewer sum rules at the $n=4, k=0$ level in Table II.

First, even before analyzing the details of the Hamiltonian for the process, the $|\Delta U_3|=\frac{1}{2}$ selection rule suggests that $\cA_{B_d /B_s \rightarrow \pi^+\pi^+K^-K^-}=0$ and $\cA_{B_d /B_s \rightarrow K^+K^+\pi^-\pi^-}=0$, suppressing momentum labels. Furthermore, the same rule also suggests that the amplitudes $\cA_{B_s\rightarrow K^+K^+\pi^-K^-} = 0$, $\cA_{B_s\rightarrow K^-\pi^-\pi^+\pi^+}=0$, $\cA_{B_d\rightarrow K^+\pi^+K^-K^-}=0$ as well as $\cA_{B_d \rightarrow \pi^+ \pi^+ \pi^- K^-}=0$. We anticipate that at leading order, where U-spin breaking is taken to be negligible ($k=0$), the decay of $B$ mesons is facilitated by an interaction Hamiltonian of the form:
\begin{equation}
    H_0 \sim c_{\frac{1}{2}}H^{(\frac{1}{2})}_{\frac{1}{2}}+ c_{-\frac{1}{2}}H^{(\frac{1}{2})}_{-\frac{1}{2}},
\end{equation}
in terms of the Wilson coefficients $c_{\pm \frac{1}{2}}$. $H_0$ facilitates the $|\Delta U_3|=\frac{1}{2}$ transitions (the superscript $(\frac{1}{2})$ is written explicitly to emphasize the transformation property under SU(2); for example, $H^{(\frac{1}{2})}_{\frac{1}{2}}$ transforms as $T^{(\frac{1}{2})}_{\frac{1}{2}}$). The above expression for $H_0$ gives us the triangle relations between amplitudes:
\begin{equation}
    0=   \cA_{B_d \rightarrow \pi^+ K^+ \pi^- K^-}-\cA_{B_d \rightarrow K^+K^+K^-K^-}-\cA_{B_d \rightarrow \pi^+\pi^+\pi^-\pi^-},
\end{equation}
\begin{equation}
    0=   \cA_{B_s \rightarrow \pi^+ K^+ \pi^- K^-}-\cA_{B_s \rightarrow K^+K^+K^-K^-}-\cA_{B_s \rightarrow \pi^+\pi^+\pi^-\pi^-},
\end{equation}
as well as:
\begin{equation}
   0= \cA_{B_s \rightarrow \pi^+\pi^+\pi^-K^-}-\cA_{B_s \rightarrow \pi^+K^+K^-K^-},
\end{equation}
\begin{equation}
    0= \cA_{B_d \rightarrow K^+K^+K^-\pi^-}-\cA_{B_d \rightarrow \pi^+K^+\pi^-\pi^-},
\end{equation}
%The two equations above immediately suggest that the decay rates for these processes satisfy:
%\begin{equation}
%    \Gamma(B_s \rightarrow \pi^+\pi^+\pi^-K^-)=\Gamma(B_s \rightarrow \pi^+K^+K^-K^-),
%\end{equation}
%\begin{equation}
%    \Gamma(B_d \rightarrow K^+K^+K^-\pi^-)=\Gamma(B_d \rightarrow \pi^+K^+\pi^-\pi^-),
%\end{equation}
with corrections occurring at $\mathcal{O}(\epsilon)$. Note that these lowest order amplitude sum rules are in agreement with \cite{gronau2013u}.

Additionally, we consider the role of the U-spin breaking operator $M_\epsilon$: the tensor product of one factor $M_\epsilon$ on $H_0$ (i.e. $k=1$) gives $M_\epsilon H_0 \sim M_\epsilon H^{(\frac{1}{2})}_{\frac{1}{2}}+M_\epsilon H^{(\frac{1}{2})}_{-\frac{1}{2}}$. The expression $M_\epsilon H^{(\frac{1}{2})}_{\pm \frac{1}{2}}$ contains terms that transform as $T^{(\frac{1}{2})}_{\pm \frac{1}{2}} \oplus T^{(\frac{3}{2})}_{\pm \frac{1}{2}}$ and so we write the new interaction Hamiltonian $H_I$ as:
\begin{equation}
    H_I \sim c_{\frac{1}{2}}H^{(\frac{1}{2})}_{\frac{1}{2}}+ c_{-\frac{1}{2}}H^{(\frac{1}{2})}_{-\frac{1}{2}}+\epsilon \Bigg( c_{\frac{1}{2}}\Big(\frac{1}{\sqrt{3}}H'^{(\frac{1}{2})}_{\frac{1}{2}}+\sqrt{\frac{2}{3}}H'^{(\frac{3}{2})}_{\frac{1}{2}}\Big)+c_{-\frac{1}{2}}\Big(-\frac{1}{\sqrt{3}}H'^{(\frac{1}{2})}_{-\frac{1}{2}}+\sqrt{\frac{2}{3}}H'^{(\frac{3}{2})}_{-\frac{1}{2}}\Big)\Bigg),
\end{equation}
where we have explicitly shown the factor of $\epsilon$ and extracted the Clebsch-Gordan coefficients.
\\
Using $H_I$, we find the following sum rules should hold to $\mathcal{O}(\epsilon)$, with corrections appearing at $\mathcal{O}(\epsilon^2)$:
\begin{equation}
\begin{aligned}
0=  \cA_{B_d \rightarrow \pi^+ K^+ \pi^- K^-}-\cA_{B_d \rightarrow K^+K^+K^-K^-}-\cA_{B_d \rightarrow \pi^+\pi^+\pi^-\pi^-} \\
+\cA_{B_s \rightarrow \pi^+\pi^+\pi^-K^-}-\cA_{B_s \rightarrow \pi^+K^+K^-K^-},\label{eq:order2_1}
\end{aligned}
\end{equation}
\begin{equation}
\begin{aligned}
0=  \cA_{B_s \rightarrow \pi^+ K^+ \pi^- K^-}-\cA_{B_s \rightarrow K^+K^+K^-K^-}-\cA_{B_s \rightarrow \pi^+\pi^+\pi^-\pi^-} \\
+\cA_{B_d \rightarrow K^+K^+K^-\pi^-}-\cA_{B_d \rightarrow \pi^+K^+\pi^-\pi^-}, \label{eq:order2_2}
\end{aligned}
\end{equation}
where we note that we get one sum rule each,  Eqs. (\ref{eq:order2_1}) and (\ref{eq:order2_2}), for each $\Delta U_3=\pm \frac{1}{2}$, as expected for $n=4$, $k=1$ in Table \ref{tab:Uspin-number-sum-rules}. Interestingly, in the end coefficients of the amplitudes turn out to be simply $\pm 1$.

\section{Conclusion}

To conclude, we have analysed various aspects of higher-order isospin and U-spin sum rules in meson decays. For $B \to n\pi$, we demonstrated that isospin sum rules exist to any order in the isospin-breaking parameter $\delta$ as long as $n$ is large enough, even when we include the effects of $\pi$--$\eta$--$\eta'$ mixing. 
\begin{comment}
We also demonstrated how $\mathcal{O}(\delta)$ sum rules for $B \to 3\pi$ at the symmetric point in phase space can be used to derive $\alpha$ to $\mathcal{O}(\delta^2)$ precision, a one-order improvement from the Gronau-London method. 
\end{comment}
Nevertheless, various issues need to be addressed before this method can be made practical, most importantly: (1) the effects of kinematic corrections on the sum rules due to meson mass splitting, and (2) insufficient statistics due to reliance on specific points in phase space.

For $B_d$ and $B_s$ decays to $n$ charged pions and/or kaons, again we demonstrated that U-spin sum rules exist to any order in U-spin breaking as long as $n$ is large enough, although the effects of kinematic corrections on these sum rules remain to be addressed. It is also worth thinking about how these U-spin sum rules might be used in the precision extraction of CKM parameters.

\section*{Acknowledgements}

The author is extremely grateful to Yuval Grossman, Wee Hao Ng, Yotam Soreq, Dean Robinson, Zoltan Ligeti, and John March-Russell for many useful conversations. The author is supported by the Prime Minister Fellowship, Prime Minister's Office, Government of Bangladesh.

%%%%%%%%%%%%%%%%%%%%%%%%%%%%%%%%%%%%%%%%%%%%%%%%%%%%%%%%%%%%%%%%%%%%%%%%%%%%%%%%%%%%%%%%%%%%%%%%%%%%
%%%%%%%%%%%%%%%%%%%%%%%%%%%%%%%%%%%%%%%%%%%%%%%%%%%%%%%%%%%%%%%%%%%%%%%%%%%%%%%%%%%%%%%%%%%%%%%%%%%%
%%%%%%%%%%%%%%%%%%%%%%%%%%%%%%%%%%%%%%%%%%%%%%%%%%%%%%%%%%%%%%%%%%%%%%%%%%%%%%%%%%%%%%%%%%%%%%%%%%%%
%%%%%%%%%%%%%%%%%%%%%%%%%%%%%%%%%%%%%%%%%%%%%%%%%%%%%%%%%%%%%%%%%%%%%%%%%%%%%%%%%%%%%%%%%%%%%%%%%%%%
%%%%%%%%%%%%%%%%%%%%%%%%%%%%%%%%%%%%%%%%%%%%%%%%%%%%%%%%%%%%%%%%%%%%%%%%%%%%%%%%%%%%%%%%%%%%%%%%%%%%

\appendix

\section{The symmetric point in phase space for $B\to n\pi$} \label{app:symmetric-point}

In general, $n\pi$ final states are made up of isospin representations from $I = 0$ to $n$, often with multiple copies for each $I$. If we want only the totally symmetric isospin representations (i.e. single copies of $I = n$, $n-2$, etc) to contribute, we require the decay amplitudes to be invariant under permutations of the charge indices, e.g. $A_{++-} = A_{+-+} = A_{-++}$. Since there is no a priori reason for the amplitudes to be permutation-invariant, this probably only occurs at specific points in phase space called the symmetric points.
One can use rotational symmetry to identify the symmetric points. Let $(c_1', \ldots, c_n') = \sigma(c_1,\ldots, c_n)$, where $\sigma \in S_n$, and $S_n$ is the permutation group of $n$ objects. If $(\pi^{c'_1}(p_1),\ldots,\pi^{c'_n}(p_n))$ can be obtained from $(\pi^{c_1}(p_1),\ldots,\pi^{c_n}(p_n))$ by a spatial rotation, then this implies that $A_{\sigma(c_1,\ldots,c_n)} = A_{c_1,\ldots,c_n}$. If this is true for all permutations $\sigma$ and for all charge compositions, then $(p_1,\ldots,p_n)$ is a symmetric point. Clearly, such a point requires that all momenta are of the same magnitude, and oriented such that the resulting geometrical figure satisfies some notion of regularity. We consider some specific examples below.

\begin{figure}[ht]
\centering
\includegraphics[width=0.8\linewidth]{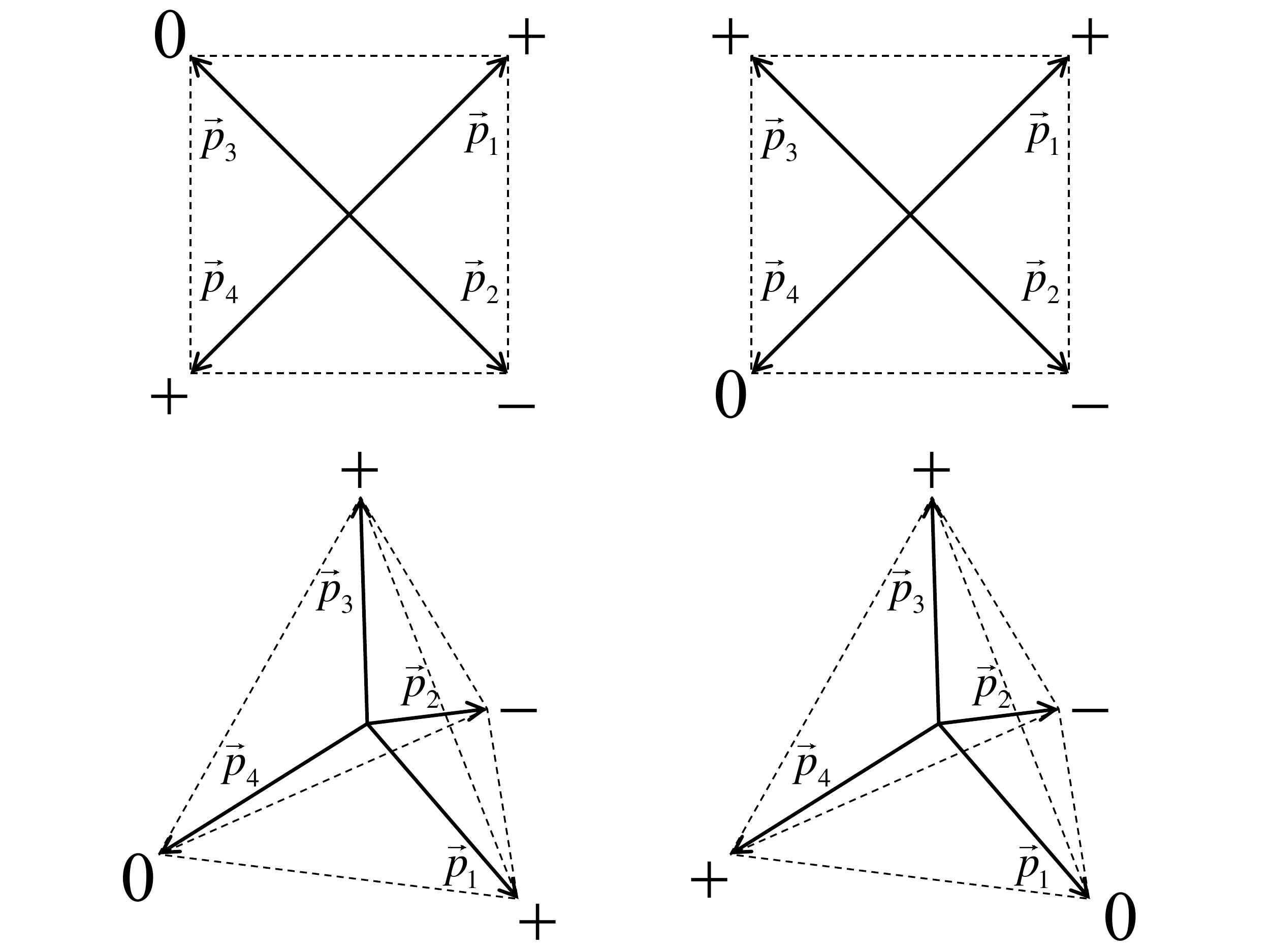}
\caption{Symmetric point for four pions. The square is not a symmetric point since there exists charge permutations that are not equivalent to rotations. For instance, the first square cannot be rotated into the second. In contrast, a regular tetrahedron is a symmetric point because any charge permutations can be achieved by a rotation. For instance, the first tetrahedron can be rotated into the second with the right choice of rotation axis.} \label{fig:4-pion-symmetric-point}
\end{figure}

For $n=2$, the entire two-pion phase space are symmetric points. For $n=3$, the symmetric points are where all three momenta form vertices of an equilateral triangle. For $n=4$, a square is not a symmetric point, because certain charge permutations of vertices cannot be obtained by rotations, see Fig.~\ref{fig:4-pion-symmetric-point}. On the other hand, a regular tetrahedron is a symmetric point because all charge permutations can be obtained by rotations. Note that the tetrahedron being a symmetric point relies on the fact that there are only three possible charges, so at least two of the vertices must carry the same charges. Were all four vertices distinctly charged, then certain permutations can only be obtained from a combination of rotations and spatial inversion. Since the weak interaction is not invariant under parity, the tetrahedron would not be a symmetric point in such a hypothetical scenario.

For $n \ge 5$, we have not been able to find any geometrical figures that has the required property. Therefore, unless other symmetry arguments can be used to force the equality of amplitudes under charge permutations, otherwise one is led to conclude that the symmetric point exists only for $n \le 4$.

\section{Conventions in the sum rules} \label{app:isospin-sum-rules-convention}

We note two differences in the convention between the sum rules here and the ones used in the original Gronau-London analysis. The first difference is that the convention given in Eq.~(\ref{eq:isospin-pion-defn}) implies that $\langle \pi^+| = -\langle 1,1|$, whereas in the Gronau-London analysis, $\langle \pi^+| = \langle 1,1|$. This accounts for any sign differences in the sum rules.

The second difference is as follows. The Gronau-London amplitude sums all distinct permutations of a charge composition, divided by the square root of the number of distinct permutations. For instance,
\begin{equation}
\begin{aligned}
&A^{GL}_{+0} = \frac{A_{+0} + A_{0+}}{\sqrt{2}}, \quad A^{GL}_{+-} = \frac{A_{+-} + A_{-+}}{\sqrt{2}}, \quad A^{GL}_{00} = A_{00},\\
&A^{GL}_{++-} = \frac{A_{++-} + A_{+-+} + A_{-++}}{\sqrt{3}}, \quad A^{GL}_{+00} = \frac{A_{+00} + A_{0+0} + A_{00+}}{\sqrt{3}}, \\
&A^{GL}_{+0-} = \frac{A_{+0-} + A_{+-0} + A_{0+-} + A_{0-+} + A_{-+0} + A_{-0+}}{\sqrt{6}}, \quad A^{GL}_{000} = A_{000}.
\end{aligned}
\end{equation}
The Gronau-London convention can be motivated by regarding all the pions as identical particles distinguished by their isospin and momentum quantum numbers, and then factorizing certain parts of the complete final state of the decaying $B$ meson as a totally symmetric pion wavefunction, multiplied by a totally symmetric spatial wavefunction, e.g. $\left(\tfrac{|+\rangle_a |0\rangle_b + |0\rangle_a |+\rangle_b}{\sqrt{2}}\right)\left(\tfrac{|p_1\rangle_a |p_2\rangle_b + |p_2\rangle_a |p_1\rangle_b}{\sqrt{2}}\right)$. In this definition, amplitudes are for transitions to these totally symmetric pion wavefunctions. Note that this factorization is actually only possible for parts of the complete final state at the symmetric point; however, for $n=2$ as used in the original Gronau-London analysis, every point in the phase space is a symmetric point.

\bibliography{B_sum_rules}

%merlin.mbs apsrev4-1.bst 2010-07-25 4.21a (PWD, AO, DPC) hacked
%Control: key (0)
%Control: author (8) initials jnrlst
%Control: editor formatted (1) identically to author
%Control: production of article title (-1) disabled
%Control: page (0) single
%Control: year (1) truncated
%Control: production of eprint (0) enabled
\begin{thebibliography}{17}%
\makeatletter
\providecommand \@ifxundefined [1]{%
 \@ifx{#1\undefined}
}%
\providecommand \@ifnum [1]{%
 \ifnum #1\expandafter \@firstoftwo
 \else \expandafter \@secondoftwo
 \fi
}%
\providecommand \@ifx [1]{%
 \ifx #1\expandafter \@firstoftwo
 \else \expandafter \@secondoftwo
 \fi
}%
\providecommand \natexlab [1]{#1}%
\providecommand \enquote  [1]{``#1''}%
\providecommand \bibnamefont  [1]{#1}%
\providecommand \bibfnamefont [1]{#1}%
\providecommand \citenamefont [1]{#1}%
\providecommand \href@noop [0]{\@secondoftwo}%
\providecommand \href [0]{\begingroup \@sanitize@url \@href}%
\providecommand \@href[1]{\@@startlink{#1}\@@href}%
\providecommand \@@href[1]{\endgroup#1\@@endlink}%
\providecommand \@sanitize@url [0]{\catcode `\\12\catcode `\$12\catcode
  `\&12\catcode `\#12\catcode `\^12\catcode `\_12\catcode `\%12\relax}%
\providecommand \@@startlink[1]{}%
\providecommand \@@endlink[0]{}%
\providecommand \url  [0]{\begingroup\@sanitize@url \@url }%
\providecommand \@url [1]{\endgroup\@href {#1}{\urlprefix }}%
\providecommand \urlprefix  [0]{URL }%
\providecommand \Eprint [0]{\href }%
\providecommand \doibase [0]{http://dx.doi.org/}%
\providecommand \selectlanguage [0]{\@gobble}%
\providecommand \bibinfo  [0]{\@secondoftwo}%
\providecommand \bibfield  [0]{\@secondoftwo}%
\providecommand \translation [1]{[#1]}%
\providecommand \BibitemOpen [0]{}%
\providecommand \bibitemStop [0]{}%
\providecommand \bibitemNoStop [0]{.\EOS\space}%
\providecommand \EOS [0]{\spacefactor3000\relax}%
\providecommand \BibitemShut  [1]{\csname bibitem#1\endcsname}%
\let\auto@bib@innerbib\@empty
%</preamble>
\bibitem [{\citenamefont {Gronau}\ and\ \citenamefont
  {London}(1990)}]{Gronau:1990ka}%
  \BibitemOpen
  \bibfield  {author} {\bibinfo {author} {\bibfnamefont {M.}~\bibnamefont
  {Gronau}}\ and\ \bibinfo {author} {\bibfnamefont {D.}~\bibnamefont
  {London}},\ }\href {\doibase 10.1103/PhysRevLett.65.3381} {\bibfield
  {journal} {\bibinfo  {journal} {Phys. Rev. Lett.}\ }\textbf {\bibinfo
  {volume} {65}},\ \bibinfo {pages} {3381} (\bibinfo {year}
  {1990})}\BibitemShut {NoStop}%
%%CITATION = PRLTA,65,3381;%%
\bibitem [{\citenamefont {Grossman}\ and\ \citenamefont
  {Robinson}(2013)}]{grossman20133}%
  \BibitemOpen
  \bibfield  {author} {\bibinfo {author} {\bibfnamefont {Y.}~\bibnamefont
  {Grossman}}\ and\ \bibinfo {author} {\bibfnamefont {D.~J.}\ \bibnamefont
  {Robinson}},\ }\href@noop {} {\bibfield  {journal} {\bibinfo  {journal}
  {Journal of High Energy Physics}\ }\textbf {\bibinfo {volume} {2013}},\
  \bibinfo {pages} {1} (\bibinfo {year} {2013})}\BibitemShut {NoStop}%
\bibitem [{\citenamefont {Gronau}(2013)}]{gronau2013u}%
  \BibitemOpen
  \bibfield  {author} {\bibinfo {author} {\bibfnamefont {M.}~\bibnamefont
  {Gronau}},\ }\href@noop {} {\bibfield  {journal} {\bibinfo  {journal}
  {Physics Letters B}\ }\textbf {\bibinfo {volume} {727}},\ \bibinfo {pages}
  {136} (\bibinfo {year} {2013})}\BibitemShut {NoStop}%
\bibitem [{\citenamefont {Gronau}(2000)}]{gronau2000u}%
  \BibitemOpen
  \bibfield  {author} {\bibinfo {author} {\bibfnamefont {M.}~\bibnamefont
  {Gronau}},\ }\href@noop {} {\bibfield  {journal} {\bibinfo  {journal}
  {Physics Letters B}\ }\textbf {\bibinfo {volume} {492}},\ \bibinfo {pages}
  {297} (\bibinfo {year} {2000})}\BibitemShut {NoStop}%
\bibitem [{\citenamefont {Grossman}\ \emph {et~al.}(2014)\citenamefont
  {Grossman}, \citenamefont {Ligeti},\ and\ \citenamefont
  {Robinson}}]{Grossman:2013lya}%
  \BibitemOpen
  \bibfield  {author} {\bibinfo {author} {\bibfnamefont {Y.}~\bibnamefont
  {Grossman}}, \bibinfo {author} {\bibfnamefont {Z.}~\bibnamefont {Ligeti}}, \
  and\ \bibinfo {author} {\bibfnamefont {D.~J.}\ \bibnamefont {Robinson}},\
  }\href {\doibase 10.1007/JHEP01(2014)066} {\bibfield  {journal} {\bibinfo
  {journal} {JHEP}\ }\textbf {\bibinfo {volume} {01}},\ \bibinfo {pages} {066}
  (\bibinfo {year} {2014})},\ \Eprint {http://arxiv.org/abs/1308.4143}
  {arXiv:1308.4143 [hep-ph]} \BibitemShut {NoStop}%
%%CITATION = ARXIV:1308.4143;%%
\bibitem [{\citenamefont {Gardner}(1999)}]{Gardner:1998gz}%
  \BibitemOpen
  \bibfield  {author} {\bibinfo {author} {\bibfnamefont {S.}~\bibnamefont
  {Gardner}},\ }\href {\doibase 10.1103/PhysRevD.59.077502} {\bibfield
  {journal} {\bibinfo  {journal} {Phys. Rev.}\ }\textbf {\bibinfo {volume}
  {D59}},\ \bibinfo {pages} {077502} (\bibinfo {year} {1999})},\ \Eprint
  {http://arxiv.org/abs/hep-ph/9806423} {arXiv:hep-ph/9806423 [hep-ph]}
  \BibitemShut {NoStop}%
%%CITATION = HEP-PH/9806423;%%
\bibitem [{\citenamefont {Gronau}\ and\ \citenamefont
  {Zupan}(2005)}]{Gronau:2005pq}%
  \BibitemOpen
  \bibfield  {author} {\bibinfo {author} {\bibfnamefont {M.}~\bibnamefont
  {Gronau}}\ and\ \bibinfo {author} {\bibfnamefont {J.}~\bibnamefont {Zupan}},\
  }\href {\doibase 10.1103/PhysRevD.71.074017} {\bibfield  {journal} {\bibinfo
  {journal} {Phys. Rev.}\ }\textbf {\bibinfo {volume} {D71}},\ \bibinfo {pages}
  {074017} (\bibinfo {year} {2005})},\ \Eprint
  {http://arxiv.org/abs/hep-ph/0502139} {arXiv:hep-ph/0502139 [hep-ph]}
  \BibitemShut {NoStop}%
%%CITATION = HEP-PH/0502139;%%
\bibitem [{\citenamefont {Schacht}\ and\ \citenamefont
  {Soni}(2021)}]{schacht2021enhancement}%
  \BibitemOpen
  \bibfield  {author} {\bibinfo {author} {\bibfnamefont {S.}~\bibnamefont
  {Schacht}}\ and\ \bibinfo {author} {\bibfnamefont {A.}~\bibnamefont {Soni}},\
  }\href@noop {} {\bibfield  {journal} {\bibinfo  {journal} {Physics Letters
  B}\ ,\ \bibinfo {pages} {136855}} (\bibinfo {year} {2021})}\BibitemShut
  {NoStop}%
\bibitem [{\citenamefont {Savage}(1991)}]{savage19913}%
  \BibitemOpen
  \bibfield  {author} {\bibinfo {author} {\bibfnamefont {M.~J.}\ \bibnamefont
  {Savage}},\ }\href@noop {} {\bibfield  {journal} {\bibinfo  {journal}
  {Physics Letters B}\ }\textbf {\bibinfo {volume} {257}},\ \bibinfo {pages}
  {414} (\bibinfo {year} {1991})}\BibitemShut {NoStop}%
\bibitem [{\citenamefont {Brod}\ \emph {et~al.}(2012)\citenamefont {Brod},
  \citenamefont {Grossman}, \citenamefont {Kagan},\ and\ \citenamefont
  {Zupan}}]{brod2012consistent}%
  \BibitemOpen
  \bibfield  {author} {\bibinfo {author} {\bibfnamefont {J.}~\bibnamefont
  {Brod}}, \bibinfo {author} {\bibfnamefont {Y.}~\bibnamefont {Grossman}},
  \bibinfo {author} {\bibfnamefont {A.~L.}\ \bibnamefont {Kagan}}, \ and\
  \bibinfo {author} {\bibfnamefont {J.}~\bibnamefont {Zupan}},\ }\href@noop {}
  {\bibfield  {journal} {\bibinfo  {journal} {Journal of High Energy Physics}\
  }\textbf {\bibinfo {volume} {2012}},\ \bibinfo {pages} {1} (\bibinfo {year}
  {2012})}\BibitemShut {NoStop}%
\bibitem [{\citenamefont {Grossman}\ and\ \citenamefont
  {Schacht}(2019)}]{grossman2019u}%
  \BibitemOpen
  \bibfield  {author} {\bibinfo {author} {\bibfnamefont {Y.}~\bibnamefont
  {Grossman}}\ and\ \bibinfo {author} {\bibfnamefont {S.}~\bibnamefont
  {Schacht}},\ }\href@noop {} {\bibfield  {journal} {\bibinfo  {journal}
  {Physical Review D}\ }\textbf {\bibinfo {volume} {99}},\ \bibinfo {pages}
  {033005} (\bibinfo {year} {2019})}\BibitemShut {NoStop}%
\bibitem [{\citenamefont {Hiller}\ \emph
  {et~al.}(2013{\natexlab{a}})\citenamefont {Hiller}, \citenamefont {Jung},\
  and\ \citenamefont {Schacht}}]{hiller2013s}%
  \BibitemOpen
  \bibfield  {author} {\bibinfo {author} {\bibfnamefont {G.}~\bibnamefont
  {Hiller}}, \bibinfo {author} {\bibfnamefont {M.}~\bibnamefont {Jung}}, \ and\
  \bibinfo {author} {\bibfnamefont {S.}~\bibnamefont {Schacht}},\ }\href@noop
  {} {\bibfield  {journal} {\bibinfo  {journal} {Physical Review D}\ }\textbf
  {\bibinfo {volume} {87}},\ \bibinfo {pages} {014024} (\bibinfo {year}
  {2013}{\natexlab{a}})}\BibitemShut {NoStop}%
\bibitem [{\citenamefont {Hiller}\ \emph
  {et~al.}(2013{\natexlab{b}})\citenamefont {Hiller}, \citenamefont {Jung},\
  and\ \citenamefont {Schacht}}]{hiller20133}%
  \BibitemOpen
  \bibfield  {author} {\bibinfo {author} {\bibfnamefont {G.}~\bibnamefont
  {Hiller}}, \bibinfo {author} {\bibfnamefont {M.}~\bibnamefont {Jung}}, \ and\
  \bibinfo {author} {\bibfnamefont {S.}~\bibnamefont {Schacht}},\ }\href@noop
  {} {\bibfield  {journal} {\bibinfo  {journal} {arXiv preprint
  arXiv:1311.3883}\ } (\bibinfo {year} {2013}{\natexlab{b}})}\BibitemShut
  {NoStop}%
\bibitem [{\citenamefont {Dery}\ \emph {et~al.}(2021)\citenamefont {Dery},
  \citenamefont {Grossman}, \citenamefont {Schacht},\ and\ \citenamefont
  {Soffer}}]{dery2021probing}%
  \BibitemOpen
  \bibfield  {author} {\bibinfo {author} {\bibfnamefont {A.}~\bibnamefont
  {Dery}}, \bibinfo {author} {\bibfnamefont {Y.}~\bibnamefont {Grossman}},
  \bibinfo {author} {\bibfnamefont {S.}~\bibnamefont {Schacht}}, \ and\
  \bibinfo {author} {\bibfnamefont {A.}~\bibnamefont {Soffer}},\ }\href@noop {}
  {\bibfield  {journal} {\bibinfo  {journal} {Journal of High Energy Physics}\
  }\textbf {\bibinfo {volume} {2021}},\ \bibinfo {pages} {1} (\bibinfo {year}
  {2021})}\BibitemShut {NoStop}%
\bibitem [{\citenamefont {Dery}\ \emph {et~al.}(2020)\citenamefont {Dery},
  \citenamefont {Ghosh}, \citenamefont {Grossman},\ and\ \citenamefont
  {Schacht}}]{Dery:2020lbc}%
  \BibitemOpen
  \bibfield  {author} {\bibinfo {author} {\bibfnamefont {A.}~\bibnamefont
  {Dery}}, \bibinfo {author} {\bibfnamefont {M.}~\bibnamefont {Ghosh}},
  \bibinfo {author} {\bibfnamefont {Y.}~\bibnamefont {Grossman}}, \ and\
  \bibinfo {author} {\bibfnamefont {S.}~\bibnamefont {Schacht}},\ }\href
  {\doibase 10.1007/JHEP03(2020)165} {\bibfield  {journal} {\bibinfo  {journal}
  {JHEP}\ }\textbf {\bibinfo {volume} {03}},\ \bibinfo {pages} {165} (\bibinfo
  {year} {2020})},\ \Eprint {http://arxiv.org/abs/2001.05397} {arXiv:2001.05397
  [hep-ph]} \BibitemShut {NoStop}%
\bibitem [{\citenamefont {Fleischer}(1999)}]{Fleischer:1999pa}%
  \BibitemOpen
  \bibfield  {author} {\bibinfo {author} {\bibfnamefont {R.}~\bibnamefont
  {Fleischer}},\ }\href {\doibase 10.1016/S0370-2693(99)00640-1} {\bibfield
  {journal} {\bibinfo  {journal} {Phys. Lett.}\ }\textbf {\bibinfo {volume}
  {B459}},\ \bibinfo {pages} {306} (\bibinfo {year} {1999})},\ \Eprint
  {http://arxiv.org/abs/hep-ph/9903456} {arXiv:hep-ph/9903456 [hep-ph]}
  \BibitemShut {NoStop}%
%%CITATION = HEP-PH/9903456;%%
\bibitem [{\citenamefont {Fleischer}(2007)}]{Fleischer:2007hj}%
  \BibitemOpen
  \bibfield  {author} {\bibinfo {author} {\bibfnamefont {R.}~\bibnamefont
  {Fleischer}},\ }\href {\doibase 10.1140/epjc/s10052-007-0391-7} {\bibfield
  {journal} {\bibinfo  {journal} {Eur. Phys. J.}\ }\textbf {\bibinfo {volume}
  {C52}},\ \bibinfo {pages} {267} (\bibinfo {year} {2007})},\ \Eprint
  {http://arxiv.org/abs/0705.1121} {arXiv:0705.1121 [hep-ph]} \BibitemShut
  {NoStop}%
%%CITATION = ARXIV:0705.1121;%%
\end{thebibliography}%
 
\end{document}